\documentclass{svmult}

\usepackage{graphicx}
\usepackage{makeidx,multicol}

\newcommand{\be}{\begin{equation}}
\newcommand{\ee}{\end{equation}}

\newcommand{\ie}{{\frenchspacing \em i.e.}}
\newcommand{\eg}{{\frenchspacing \em e.g.}}

\newcommand{\x}{{\vec{x}}}

\newcommand{\A}{{\tens{A}}}
\newcommand{\B}{{\tens{B}}}
\newcommand{\C}{{\tens{C}}}

\newcommand{\DD}{{\tens{D}}}
\newcommand{\F}{{\tens{F}}}
\newcommand{\G}{{\tens{G}}}
\newcommand{\HH}{{\tens{H}}}
\newcommand{\II}{{\tens{I}}}
\newcommand{\R}{{\tens{R}}}

\newcommand{\M}{{\tens{M}}}

\newcommand{\bk}{{\bf k}}

\newcommand{\bx}{{\bf x}}
\newcommand{\bu}{{\bf u}}
\newcommand{\Tr}{{\rm Tr}}

\font\bfmath=cmmib10
\def\vth{
\hbox{\bfmath\char'22}}    
\def\vmu{\hbox{\vec{\mu}}}    
\def\bLambda{
\hbox{\bfmath\char'0003}}   

\def\vpsi{\bfmath\char'11}

\def\beq#1{\begin{equation}\label{#1}}
\def\eeq{\end{equation}}
\def\eq#1{equation~(\ref{#1})}

\def\vthml{\vth_{\hbox{ML}}}
\def\expec#1{\langle#1\rangle}
\def\bigexpec#1{\left\langle#1\right\rangle}
\def\Ell{{\cal L}}
\def\crr{\cr\noalign{\vskip 4pt}}

\title*{Statistical techniques in cosmology}
\author{Alan Heavens}
\institute{Institute for Astronomy, University of Edinburgh,
Blackford Hill, Edinburgh EH9 3HJ, afh@roe.ac.uk\\Lectures given at `Francesco Lucchin' Summer School, Bertinoro, May, 2009, and SISSA, Italy, May 2010}

\begin{document}

\maketitle

\section{Introduction}


In these lectures I cover a number of topics in cosmological data analysis.  I concentrate on general techniques which are common in cosmology, or techniques which have been developed in a cosmological context.  In fact they have very general applicability, for problems in which the data are interpreted in the context of a theoretical model, and thus lend themselves to a Bayesian treatment.

We consider the general problem of estimating parameters from
data, and consider how one can use Fisher matrices to analyse
survey designs before any data are taken, to see whether the
survey will actually do what is required.  We outline numerical methods for estimating parameters from data, including Monte Carlo Markov Chains and the Hamiltonian Monte Carlo method.  We also look at Model Selection, which covers various scenarios such as whether an extra parameter is preferred by the data, or  answering wider questions such as which theoretical framework is favoured, using General Relativity and braneworld gravity as an example.   These notes are not a literature review, so there are relatively few references.  Some of the derivations follow the excellent notes of Licia Verde \cite{VerdeNotes} and Andrew Hamilton \cite{Hamilton05}.

After this introduction, the sections are:
\begin{itemize}
\item{Parameter Estimation}
\item{Fisher Matrix analysis}
\item{Numerical methods for Parameter Estimation}
\item{Model Selection}
\end{itemize}

\subsection{Notations:}
\begin{itemize}
\item{{\em Data} will be called $\bx$, or $x$, or $x_i$, and is written as a vector, even if it is a 2D image.}
\item{{\em Model parameters} will be called $\vth$, or $\theta$ or $\theta_\alpha$}
\end{itemize}

\subsection{Inverse problems}

Most data analysis problems are {\em inverse problems}.  You have
a set of data $\x$, and you wish to interpret the data in some way.  Typical classifications are:
\begin{itemize}
\item{Hypothesis testing}
\item{Parameter estimation}
\item{Model selection}
\end{itemize}
Cosmological examples of the first type include 
\begin{itemize}
\item{Are CMB data consistent with the hypothesis that the initial fluctuations were gaussian, as predicted (more-or-less) by the simplest inflation theories?}
\item{Are large-scale structure observations consistent with the hypothesis that the Universe is spatially flat?}
\end{itemize}
Cosmological examples of the second type include
\begin{itemize}
\item{In the Big Bang model, what is the value of the matter density parameter?}
\item{What is the value of the Hubble constant?}
\end{itemize}
Model selection can include slightly different types (but are mostly concerned with larger, often more qualitative questions):
\begin{itemize}
\item{Do cosmological data favour the Big Bang theory or the Steady State theory?}
\item{Is the gravity law General Relativity or higher-dimensional?}
\item{Is there evidence for a non-flat Universe?}
\end{itemize}
Note that the notion of a {\em model} can be a completely different paradigm (the first two examples), or basically the same model, but with a different parameter set.  In the third example, we are comparing a flat Big Bang model with a non-flat one.  The latter has an additional parameter, and is considered to be a different model.  Similar problems exist elsewhere in astrophysics, such as how many absorption-line systems are required to account for an observed feature? 

These lectures will principally be concerned with questions of the last two types, {\em parameter estimation} and {\em model selection}, but will also touch on experimental design and error forecasting.  Hypothesis testing can be treated in a similar manner.

\section{Parameter estimation}

We collect some data, and wish to interpret them in terms of a {\em model}.  A model is a theoretical framework which we assume is true.  It will typically have some parameters $\vth$ in it, which you want to determine.  The goal of parameter estimation is to provide estimates of the parameters, and their errors, or ideally the whole
probability distribution of $\vth$, given the data  \x.  This is called the {\em posterior probability distribution} i.e. it is the probability that the parameter takes certain values, {\em after} doing the experiment (as well as assuming a lot of other things):
\begin{equation}
p(\vth|\x).
\end{equation}
This is an example of RULE 1\footnote{There is no rule $n$: $n>1$.}: Start by thinking about what it is you want to know, and write it down mathematically. 

From $p(\vth|\x)$ one can calculate the expectation values of the
parameters, and their errors.    Note that we are immediately taking a Bayesian view of probability, as a {'em degree of belief}, rather than a frequency of occurrence in a set of trials.

\subsection{Forward modelling}

Often, what may be easily calculable is
not this, rather the opposite, $p(\x|\vth)$\footnote{If you are confused about $p(A|B)$ and $p(B|A)$ consider if A=pregnant and B=female.  $p(A|B)$ is a few percent, $p(B|A)$ is unity}.   The opposite is sometimes referred to as {\em forward modelling} - i.e. if we know what the parameters are, we can compute the expected distribution of the data.  Examples of forward modelling distributions include the common ones - binomial,  Poisson, gaussian etc. or may be more complex, such as the predictions for the CMB power spectrum as a function of cosmological parameters.  As a concrete example, consider
a model which is a gaussian with mean $\mu$ and variance
$\sigma^2$.   The model has two parameters, $\vth=(\mu,\sigma)$, and the probability of a single variable $x$ given the
parameters is
\begin{equation}
p(x|\vth)={1\over \sqrt{2\pi}\sigma} \exp\left[-{(x-\mu)^2\over
2\sigma^2}\right],
\end{equation}
but this is not what we actually want.  However, we can relate
this to $p(\vth|x)$ using Bayes' Theorem, here written for a more general data vector $\x$:
\begin{equation}
p(\vth|\x) = {p(\vth,\x)\over p(\x)} = {p(\x|\vth)p(\vth)\over
p(\x)}.
\label{Bayes}
\end{equation}
\begin{itemize}
\item{$p(\vth|\x)$ is the {\em posterior} probability for the parameters.}
\item {$p(\x|\vth)$ is called the {\em Likelihood} and given its own symbol $L(\x;\vth)$.}
\item $p(\vth)$ is called the {\em prior}, and expresses what we know about the
parameters prior to the experiment being done.  This may be the
result of previous experiments, or theory (e.g. some parameters,
such as the age of the Universe, may have to be positive). In the absence of
any previous information, the prior is often assumed to be a constant (a `flat prior').
\item $p(\x)$ is the {\em evidence}.
\end{itemize}
For parameter estimation, the evidence simply acts to normalise the probabilities, 
\begin{equation}
p(\x) = \int d\vth\, p(\x|\vth)\,p(\vth)
\label{evidence}
\end{equation}
and the {\em relative} probabilities of the parameters do not depend on it, so it is often ignored and not even calculated.

However, the evidence does play an important role in {\em model selection}, when more than one theoretical model is being considered, and one wants to choose
which model is most likely, whatever the parameters are. We turn to this later.

Actually all the probabilities above should be conditional probabilities, given any prior information $I$ which we may have.  For clarity, I have omitted these for now.  $I$ may be the result of previous experiments, or may be a theoretical prior, in the absence of any data.  In such cases, it is common to adopt the {\em principle of indifference} and assume that all values of the parameter(s) is (are) equally likely, and take $p(\vth)$=constant (perhaps within some finite bounds, or if infinite bounds, set it to some arbitrary constant and work with an unnormalised prior).   This is referred to as a {\em flat prior}.  Other choices can be justified.

Thus for flat priors, we have simply
\begin{equation}
p(\vth|\x) \propto L(\x;\vth).
\end{equation}
Although we may have the full probability distribution for the parameters, often one simply
uses the peak of the distribution as the estimate of the parameters.  This is then a {\em Maximum
Likelihood} estimate.  Note that if the priors are not flat, the peak in the posterior $p(\vth|\x)$ is not necessarily the
maximum likelihood estimate.

A `rule of thumb' is that if the priors are assigned theoretically,
and they influence the result significantly, the data are usually
not good enough.  (If the priors come from previous experiment, the situation is different - we can be more certain 
that we really have some prior knowledge in this case).

Finally, note that this method does not generally give a goodness-of-fit, only relative probabilities.  It is still common to compute $\chi^2$ at this point to check the fit is sensible.

\subsection{Updating the probability distribution for a parameter}

One will often see in the literature forecasts for a new survey, where it is assumed that we will know quite a lot about cosmological parameters from another experiment.  Typically these days it is {\em Planck}, which is predicted to constrain many cosmological parameters very accurately.  Often people `include a Planck prior'.  What does this mean, and is it justified?  Essentially,  what is assumed is that by the time of the survey, Planck will have happened, and we can combine results.  We can do this in two ways: regard Planck+survey as new data, or regard the survey as the new data, but our prior information has been set by what we know from Planck.  If Bayesian statistics makes sense, it should not matter which we choose.  We show this now.

If we obtain some more information, from a new
experiment, then we can use Bayes' theorem to update our estimate of
the probabilities associated with each parameter. The problem
reduces to that of showing that adding the results of a new
experiment to the probability of the parameters is the same as doing
the two experiments first, and then seeing how they both affect the
probability of the parameters. In other words it should not matter
how we gain our information, the effect on the probability of the
parameters should be the same.

We start with Bayes' expression for the posterior probability of a
parameter (or more generally of some hypothesis), where we put explicitly that all probabilities are conditional on some prior information $I$.
\be
    p(\vth| \bx I) = \frac{p(\vth|I) p(\bx |\vth I)}{p(\bx |I)}.
\label{update}
\ee
Let say we do a new experiment with new data, $\bx '$.  We have two ways to analyse the new data:
\begin{itemize}
\item{Interpretation 1:  we regard $\bx '$ as the dataset, and $\bx I$ (means $\bx $ {\em and} $I$) as the new prior information.}
\item{Interpretation 2: we put all the data together, and call it $\bx '\bx $, and interpret it with the old prior information $I$.}
\end{itemize}

If Bayesian inference is to be consistent, it should not matter which we do.

Let us start with interpretation 1. We rewrite Bayes' theorem, equation (\ref{update}) by changing datasets $\bx  \rightarrow \bx '$, 
and letting the old data become part of the prior information $I \rightarrow I' = \bx I$.
Bayes' theorem is now
\be
    p(\vth| \bx' I') = \frac{p(\vth|\bx I) p(\bx '|\vth \bx I)}{p(\bx '|\bx I)}.
\ee
We now notice that the new prior in this expression is just the old posteriori
probability from equation (\ref{update}), and that the new likelihood is
just
\be
    p(\bx '|\bx \vth I) = \frac{p(\bx '\bx |\vth I)}{p(\bx |\vth I)}.
\ee
Substituting this expression for the
new likelihood:
\be
p(\vth| \bx I') = \frac{p(\vth|\bx I) p(\bx '\bx |\vth I)}{p(\bx '|\bx I)p(\bx |\vth I)}.
\ee
Using Bayes' theorem again on the first term on the top and the second on the bottom,
we find
\be
p(\vth| \bx  I') = \frac{p(\vth|I) p(\bx '\bx |\vth I)}{p(\bx '|\bx I)p(\bx |I)},
\ee
and simplifying the bottom gives finally
\be
    p(\vth|\bx  I') = \frac{p(\vth|I) p(\bx '\bx |\vth I)}{p(\bx '\bx |I)}=p(\vth|([\bx\bx '] I)
\ee
which is Bayes' theorem in Interpretation 2.  i.e. it has the same form as equation (\ref{update}), the outcome from the
initial experiment, but now with the data $\bx $ replaced by $\bx '\bx $.  In other words, 
we have shown that $\bx \rightarrow \bx '$ and $I \rightarrow \bx I$ is equivalent to
$\bx \rightarrow \bx '\bx $. This shows us that it doesn't matter how we
add in new information.  Bayes' theorem gives us a natural way of improving our
statistical inferences as our state of knowledge increases.

\subsection{Errors}

Let us assume we have a posterior probability distribution, which is single-peaked.  Two common estimators (indicated by a hat: $\hat{\vth}$) of the parameters are the peak (most probable) values, or the mean,
\begin{equation}
\hat{\vth} = \int d\vth\,\vth\,p(\vth|\x).
\end{equation}
An estimator is {\em unbiased} if its expectation value is the true value $\vth_0$: 
\begin{equation}
\langle \hat{\vth} \rangle = \vth_0.
\end{equation}
Let us assume for now that the prior is flat, so the posterior is proportional to the likelihood.  This can be relaxed. 
Close to the peak, a Taylor expansion of the log likelihood implies that locally it is a mutivariate gaussian {\em in parameter space}:
\begin{equation}
\ln L(\x;\vth) = \ln L(\x;\vth_0) + \frac{1}{2}(\vth_\alpha-\vth_{0\alpha})\frac{\partial^2 \ln L}{\partial \vth_\alpha \partial \vth_\beta}(\vth_\beta-\vth_{0\beta}) + \ldots
\end{equation}
or
\be
L(\x;\vth) = L(\x;\vth_0) \exp\left[-\frac{1}{2}(\vth_\alpha-\vth_{0\alpha})H_{\alpha\beta}(\vth_\beta-\vth_{0\beta})\right].
\ee 
The Hessian matrix $\HH_{\alpha\beta} \equiv -\frac{\partial^2 \ln L}{\partial \theta_\alpha \partial \theta_\beta}$ controls whether the estimates of $\theta_\alpha$ and $\theta_\beta$ are correlated or not.  If it is diagonal, the estimates are uncorrelated.  Note that this is a statement about {\em estimates} of the quantities,  not the quantities themselves, which may be entirely independent, but if they have a similar effect on the data, their estimates may be correlated.  Note that in cases of practical interest, the likelihood may not be well described by a multivariate gaussian at levels which set the interesting credibility levels (e.g. 68\%).  We turn later to how to proceed in such cases.
%

\subsection{Conditional and marginal errors}

If we fix all the parameters except one, then the error is given by the curvature along a line through the likelihood (posterior, if prior is not flat):
\be
\sigma_{\rm conditional,\alpha} = \frac{1}{\sqrt{\HH_{\alpha\alpha}}}.
\ee
This is called the {\em conditional
error}, and is the minimum error bar attainable on $\theta_\alpha$ if
all the other parameters are known. {\em It is rarely relevant and
should almost never be quoted.}

\subsection{Marginalising over a gaussian likelihood}

The marginal distribution of $\vth_1$ is obtained by integrating over the other parameters:
\be
p(\vth_1) = \int d\vth_2 \ldots d\vth_N p(\vth)
\ee
a process which is called {\em marginalisation}.  Often one sees marginal distributions of all parameters in pairs, as a way to present some complex results.  In this case 
two variables are left out of the integration.

If you plot such error ellipses, you {\em must} say what contours you plot.  If you say you plot $1\sigma$ and $2\sigma$ contours, I don't know whether this is for the joint distribution (i.e. 68\% of the probability lies within the inner contour), or whether $68\%$ of the probability of a single parameter lies within the bounds projected onto a parameter axis.  The latter is a $1\sigma$, single-parameter error contour (and corresponds to $\Delta\chi^2=1$), whereas the former is a $1\sigma$ contour for the joint distribution, and corresponds to $\Delta\chi^2=2.3$.   

Note that $\Delta\chi^2 = \chi^2 - \chi^2(minimum)$, where
\be
\chi^2 = \sum_{i}\frac{(x_i-\mu_i)^2}{\sigma_i^2}
\ee
for data $x_i$ with $\mu_i=\langle x_i\rangle$ and variance $\sigma_i^2$.  If the data are correlated, this generalises to
\be
\chi^2 = \sum_{ij}(x_i-\mu_i)C_{ij}^{-1}(x_j-\mu_j)
\ee
where
$C_{ij}=\langle (x_i-\mu_i)(x_j-\mu_j)\rangle$.

For other dimensions, see Table 1, or read...

\subsubsection{The Numerical Recipes bible, chapter 15.6 \cite{NumRec}}

Read it.  Then, when you need to plot some error contours,  read it again.

\begin{table}
\newpage
\caption{$\Delta \chi^2$ for  joint parameter estimation for 1, 2 and 3 parameters.}
\label{table:chisq}
\begin{tabular}{|c|c|c|c|c|}
\hline
$\sigma$&p&M=1&M=2&M=3\\
\hline
1$\sigma$&68.3\%&1.00&2.30&3.53\\
2$\sigma$&95.4\%&4.00&6.17&8.02\\
3$\sigma$&99.73\%&9.00&11.8&14.2\\
\hline
\end{tabular}
\end{table}

Note that some of the results I give assume the likelihood (or posterior) is well-approximated by a multivariate gaussian.  This may not be so. If your posterior is a single peak, but is not well-approximated by a multivariate gaussian, label your contours with the enclosed probability.  If the likelihood is complicated (e.g. multimodal), then you may have to plot it and leave it at that - reducing it to a maximum likelihood point and error matrix is not very helpful.  Not that in this case, the mean of the posterior may be unhelpful - it may lie in a region of parameter space with a very small posterior.

A multivariate gaussian likelihood is a common assumption, so it is useful to compute marginal errors for this rather general situation.  The simple result is that the marginal error on parameter $\theta_\alpha$ is
\begin{equation}
\sigma_\alpha = \sqrt{(\HH^{-1})_{\alpha\alpha}}.
\end{equation}
Note that we invert the Hessian matrix, and then take the square root of the diagonal components.  Let us prove this important result.  In practice it is often used to estimate errors for a future experiment, where we deal with the expectation value of the Hessian, called the {\em Fisher Matrix}:
\begin{equation}
\F_{\alpha\beta} \equiv \langle\HH_{\alpha\beta}\rangle = \left\langle -\frac{\partial^2 \ln L}{\partial \vth_\alpha \partial \vth_\beta} \right\rangle.
\end{equation}
We will have much more to say about Fisher matrices later.   The expected error on $\theta_\alpha$ is thus
\begin{equation}
\sigma_\alpha = \sqrt{(\F^{-1})_{\alpha\alpha}}.
\label{marg}
\end{equation}
It is always at least as large as the expected conditional error.  Note: this result applies for gaussian-shaped likelihoods, and is useful for experimental design.  For real data, you would do the marginalisation a different way - see later.

To prove this, we will use characteristic functions.

\subsubsection{Characteristic functions}
 In probability theory the Fourier Transform of a probability distribution function
 is known as the {\em characteristic function}.  For a multivariate distribution with $N$ parameters, it is defined by
 \begin{equation}
    \phi(\bk) = \int \! d^N\vth\, p(\vth) e^{-i \bk\cdot\vth}
 \end{equation}
with reciprocal relation
 \begin{equation}
    p(\vth) =\int \!\frac{d^N\bk}{(2 \pi)^N} \, \phi(\bk) e^{i\bk\cdot\vth}
 \end{equation}
 (note the choice of where to put the factors of $2 \pi$ is not universal).
Hence the characteristic function is also the expectation value of
$e^{-i \bk\cdot\vth}$:
 \begin{equation}
        \phi(\bk) = \langle e^{-i\bk.\vth}\rangle.
 \end{equation}

Part of the power of characteristic functions is the ease with
which one can generate all of the moments of the distribution by
differentiation:
\begin{equation}
    \langle \vth_\alpha^{n_\alpha}\ldots \vth_\beta^{n_\beta}\rangle = \left[ \frac{\partial^{n_\alpha+\ldots+n_\beta}\phi(\bk)}{\partial(-i \bk_\alpha)^{n_\alpha}\ldots \partial(-i \bk_\beta)^{n_\beta}}\right]_{\bk={\bf 0}} .
\end{equation}
 This can be seen if one expands $\phi(\bk)$ in a
power series, using
\begin{equation}
\exp(\alpha) = \sum_{i=0}^\infty \frac{\alpha^n}{n!},
\end{equation}
giving
\begin{equation}
    \phi(\bk) = 1 - i \bk \cdot \langle \vth \rangle - \frac{1}{2} \sum_{\alpha\beta} \bk_\alpha\bk_\beta  \langle \vth_\alpha \vth_\beta \rangle + \ldots.
\end{equation}
Hence for example we can compute the mean 
\begin{equation}
\langle\vth_\alpha\rangle = \left[ \frac{\partial\phi(\bk)}{\partial(-i \bk_\alpha)}\right]_{\bk={\bf 0}}
\end{equation}
and the covariances, from
 \begin{equation}
\langle\vth_\alpha\vth_\beta\rangle = \left[ \frac{\partial^2\phi(\bk)}{\partial(-i \bk_\alpha)\partial(-i\bk_\beta)}\right]_{\bk={\bf 0}}.
\label{cov}
\end{equation}
(Putting $\alpha=\beta$ yields the variance of $\vth_\alpha$ after subtracting the square of the mean).

\subsection{The expected marginal error on $\vth_\alpha$ is $\sqrt{(\F^{-1})_{\alpha\alpha}}$}
The likelihood is here assumed to be a multivariate gaussian, with expected hessian given by the Fisher matrix.  Thus (suppressing ensemble averages)
\begin{equation}
L(\vth) = \frac{1}{(2\pi)^{M/2} \sqrt{\det{\F}}} \exp\left(-\frac{1}{2} \vth^T \F \vth\right),
\end{equation}
where $T$ indicates transpose, and for simplicity I have assumed the parameters have zero mean (if not, just redefine $\vth$ as the difference between $\vth$ and the mean).  We proceed by diagonalising the quadratic, then computing the characteristic function, and compute the covariances using equation (\ref{cov}). This is achieved in the standard way by rotating the parameter axes:
\begin{equation}
\vpsi = \R \vth
\end{equation}
for a matrix $\R$.  Since $\F$ is real and symmetric, $\R$ is orthogonal, $\R^{-1} = \R^T$.  Diagonalising gives 
\begin{equation}
\vth^T \F \vth = \vpsi^T \R \F \R^T \vpsi,
\end{equation}
and the diagonal matrix composed of the eigenvalues of $\F$ 
\begin{equation}
\bLambda=\R \F \R^T,
\end{equation}
Note that the eigenvalues of $\F$ are positive, as $\F$ must be positive-definite.

The characteristic function is
\begin{equation}
\phi(\bk) = \frac{1}{(2\pi)^{M/2} \sqrt{\det{\F}}} \int d^M \vpsi \exp\left(-\frac{1}{2} \vpsi^T \bLambda  \F \vpsi\right) \exp(-i\bk^T \R^T\vpsi)
\end{equation}
where we exploit the fact that the rotation has unit Jacobian to change $d^M\vth$ to $d^M\vpsi$.  If we define ${\bf K} \equiv \R \bk$, 
\begin{equation}
\phi(\bk) = \frac{1}{(2\pi)^{M/2} \sqrt{\det{\F}}} \int d^M \vpsi \exp\left(-\frac{1}{2} \vpsi^T \bLambda  \vpsi\right) \exp(-i{\bf K}^T\vpsi)
\end{equation}
and since $\bLambda$ is diagonal, the first exponential is a sum of squares, which we can integrate separately, using 
\begin{equation}
\int_{-\infty}^\infty d\psi \exp(-\Lambda\psi^2/2)\exp(-iK\psi) =  \sqrt{2\pi/\Lambda}\exp[-K^2/(2\Lambda)].
\end{equation}   
All multiplicative factors cancel (since the rotation preserves the eigenvalues, so $\det(\F) = \prod \bLambda_\alpha$), and we obtain
\begin{equation}
\phi(\bk) = \exp\left(-\sum_i K_i^2/(2\Lambda_i)\right) = \exp\left(-\frac{1}{2} {\bf K}^T \bLambda^{-1}{\bf K}\right) = \exp\left(-\frac{1}{2} {\bk}^T \F^{-1}{\bk}\right)
\end{equation}
where the last result follows from $ {\bf K}^T \bLambda^{-1}{\bf K} = \bk^T (\R^T \bLambda^{-1} \R) \bk = \bk^T \F^{-1}\bk$.

Having obtained the characteristic function, the result (\ref{marg}) follows immediately from equation (\ref{cov}).

\subsection{Marginalising over `amplitude' variables}

It is not uncommon to want to marginalise over a nuisance parameter which is a simple scaling variable.  Examples include a calibration uncertainty, or perhaps an unknown gain in an electronic detector.  This can be done analytically for a gaussian data space:
\be
L(\vth;\bx) = \frac{1}{(2\pi)^{N/2}\sqrt{\det \C}} \exp\left[-\frac{1}{2}\sum_{ij}(\bx_i-\bx_i^{th})\C_{ij}^{-1}(\bx_j-\bx_j^{th})\right]
\ee
where we have $N$ data points with covariance matrix $\C_{ij}\equiv \langle(\bx_i-\bx_i^{th})(\bx_j-\bx_j^{th})\rangle$, and the model data values $\bx_i^{th}$ and $\C$ depend on the parameters.  Let us now assume that the covariance matrix does not depend on the parameters, so the parameter dependence is only through $\bx^{th}(\vth)$, and 
further we assume that one of the parameters simply scales the theoretical signal. i.e. $\bx^{th}(\vth)=A \bx^{th}(\vth|A=1)$.   If we want the likelihood of all the other parameters (call this $L'(\vth';\bx)$, with one fewer parameter), marginalised over the unknown $A$, then we can integrate:
\begin{eqnarray}
L(\vth';\bx) &=& \int dA L(\vth;\bx) p(A) \nonumber\\
&=& \frac{1}{(2\pi)^{N/2}\sqrt{\det \C}}\int dA  \exp\left[-\frac{1}{2}\sum_{ij}(\bx_i-A\bx_i^{th})\C_{ij}^{-1}(\bx_j-A\bx_j^{th})\right] p(A)
\end{eqnarray}
where $p(A)$ is the prior for $A$.  If we take a uniform (unnormalised!) prior on $A$ between limits $\pm\infty$, and the theoretical $\bx$ are now at $A=1$, then we can integrate the quadratic.  You can do this.

Exercise: do this.

\section{Fisher Matrix Analysis}\label{FisherSec}

This has been adapted from \cite{TTH} (hereafter TTH).

How accurately can we estimate model parameters from a given data
set? This question was basically answered 60 years ago
\cite{Fisher}, and we will now summarize the results, which are
both simple and useful.

Suppose for definiteness that our data set consists of $N$ real
numbers $x_1,\>x_2,...,x_N$, which we arrange in an
$N$-dimensional vector $\x$. These numbers could for instance
denote the measured temperatures in the $N$ pixels of a CMB sky
map, the counts-in-cells of a galaxy redshift survey, 
$N$ coefficients of a Fourier expansion of an observed
galaxy density field, or the number of gamma-ray bursts observed
in $N$ different flux bins. Before collecting the data, we think
of $\x$ as a random variable with some probability distribution
$L(\x;\vth)$, which depends in some known way on a vector of $M$
model parameters $\vth= (\theta_1, \theta_2, ..., \theta_M)$.

Such model parameters might for instance be the spectral index of
density fluctuations, the Hubble constant $h$, the cosmic density
parameter $\Omega$ or the mean redshift of gamma-ray bursts. We
will let $\vth_0$ denote the true parameter values and let $\vth$
refer to our estimate of $\vth$. Since $\vth$ is some function of
the data vector $\x$, it too is a random variable. For it to be a
good estimate, we would of course like it to be unbiased, {\ie},
\begin{equation}
\label{BiasEq}
\expec{\vth} = \vth_0,
\end{equation}
and give as small error bars as possible, {\ie}, minimize the
standard deviations
\begin{equation}
\label{SdevEq}
\Delta\theta_\alpha\equiv\left(\bigexpec{\theta_\alpha^2}-\expec{\theta_\alpha}^2\right)^{1/2}.
\end{equation}
In statistics jargon, we want the BUE $\theta_\alpha$, which
stands for the ``Best Unbiased Estimator".

A key quantity in this context is the so-called {\it Fisher
information matrix}, defined as
\begin{equation}
\label{FisherDefEq}
\F_{\alpha\beta} \equiv
\bigexpec{{\partial^2\Ell\over\partial\theta_\alpha\partial\theta_\beta}},
\end{equation}
where 
\begin{equation}
\Ell\equiv -\ln L. 
\end{equation}
Another key quantity is the {\it
maximum likelihood estimator}, or {\it ML-estimator} for brevity,
defined as the parameter vector $\vthml$ that maximizes the
likelihood function $L(\x;\vth)$.

Using this notation, a number of powerful theorems have been
proven (see {\eg} \cite{KK},\cite{KS}):
\begin{enumerate}

\item For any unbiased estimator, $\Delta\theta_\alpha \geq
1/\sqrt{\F_{\alpha\alpha}}$ (the {\em Cram\' er-Rao} inequality).


\item If an unbiased estimator attaining (``saturating'' ) the Cram\' er-Rao bound exists, it is the ML estimator (or a function thereof). 

\item The ML-estimator is asymptotically BUE.

\end{enumerate}
The first of these theorems
thus places a firm lower limit on the error bars that one can
attain, regardless of which method one is using to estimate the
parameters from the data.   You won't do better, but you might do worse.

The normal case is that the other parameters are estimated from
the data as well, in which case, as we have seen, the minimum standard deviation
rises to
\begin{equation}
\Delta\theta_\alpha \geq (\F^{-1})_{\alpha\alpha}^{1/2}.
\end{equation}
This is called the {\em marginal error}, and I reemphasise that this is normally the
relevant error to quote.

The second theorem shows that maximum-likelihood (ML) estimates
have quite a special status: if there is a best method, then the
ML-method is the one. Finally, the third result basically tells us
that in the limit of a very large data set, the ML-estimate for
all practical purposes is the best estimate, the one that for
which the Cram\'er-Rao inequality becomes an equality\footnote{This is sometimes called 'saturating the Cram\'er-Rao bound'}. It is these
nice properties that have made ML-estimators so popular.

Note that conditional and marginal errors coincide if \F\ is
diagonal.  If it is not, then the {\em estimates} of the
parameters are correlated (even if the parameters themselves are
uncorrelated).  e.g. in the example shown in Fig. 1, estimates of
the baryon density parameter $\Omega_b$ and the dark energy
equation of state $w\equiv p/\rho c^2$ are strongly correlated
with WMAP data alone, so we will tend to overestimate both
$\Omega_b$ and $w$, or underestimate both. However, the value of
$\Omega_b$ in the Universe has nothing obvious to do with $w$ -
they are independent.
\begin{figure}
\centering
\includegraphics[height=8cm]{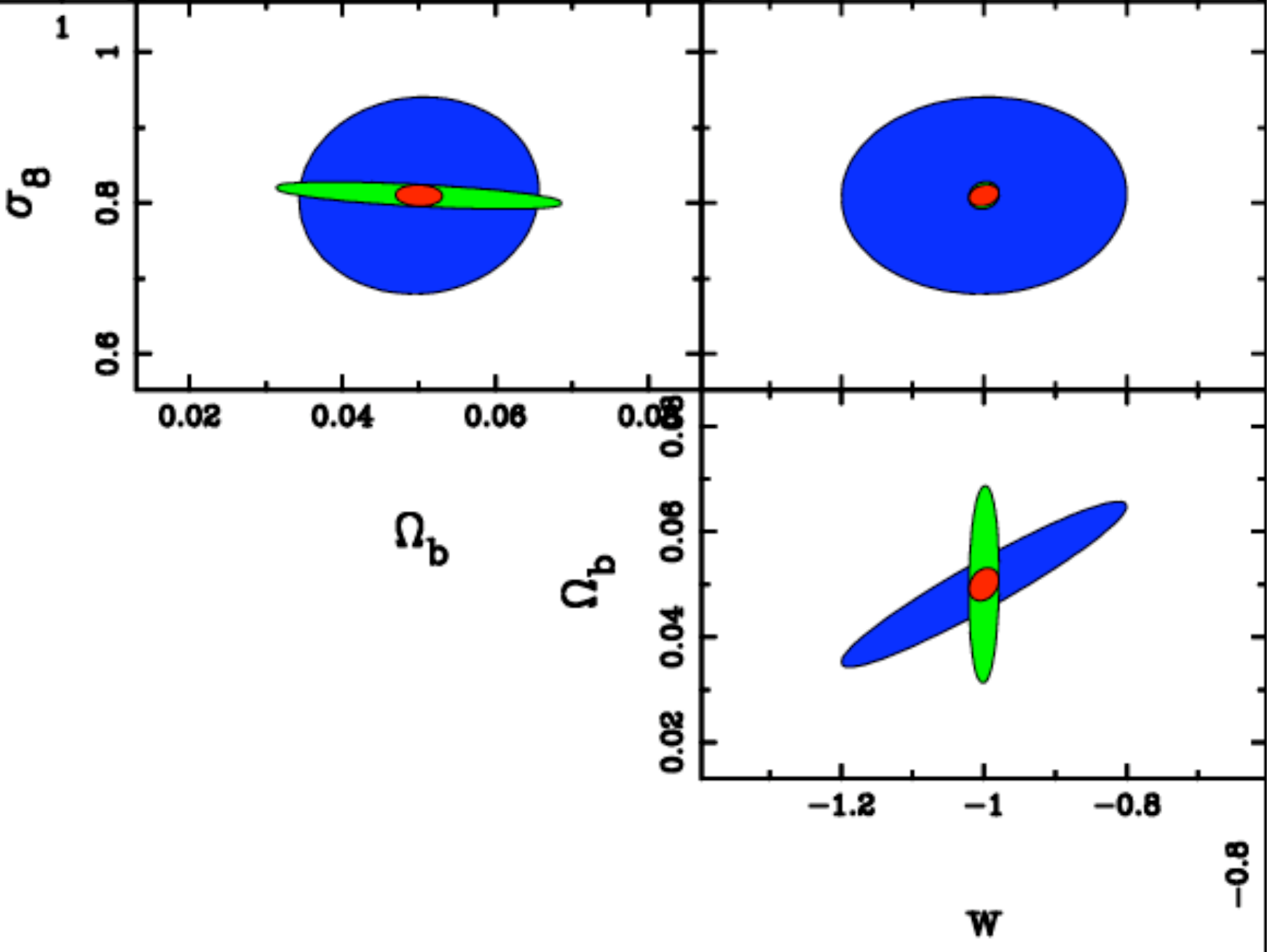}
%
%
\caption{The expected error ellipses for cosmological parameters
  ($\sigma_8$, baryon density parameter $\Omega_b$, and
  dark energy equation of state $w \equiv p/\rho c^2$)
  from a 3D weak lensing survey of 1000 square degrees, with a median redshift
  of 1 and a photometric redshift error of 0.15.  Probabilities are
  marginalised over all other parameters, except that $n=1$ and a flat
  Universe are assumed. Dark ellipses represent a prior from WMAP,
  pale represents the 3D lensing survey alone, and the central ellipses show the
  combination (from Kitching T., priv. comm.).}
\label{VISTA}       
\end{figure}

\subsection{Cram\'er-Rao inequality: proof for simple case}

This discussion follows closely Andrew Hamilton's lectures to the Valencia Summer School in 2005 \cite{Hamilton05}.

At the root of the Cram\'er-Rao inequality is the Schwarz inequality.  If we have estimators for two parameters $\theta_1$ and $\theta_2$, then it is clear that the expectation value of
\begin{equation}
\left\langle \left(\Delta\hat\theta_1 + \lambda\Delta\hat\theta_2\right)^2\right\rangle \ge 0
\label{SI}
\end{equation}
where $\Delta\hat\theta_\alpha\equiv \hat\theta_\alpha-\theta_{0\alpha}$ is the difference between the estimate and the true value.  Evidently equation \ref{SI} holds for any $\lambda$.  It is easy to show that the left-hand-side is a minimum if we choose $\lambda = -\langle\Delta\hat\theta_1 \Delta\hat\theta_2\rangle/\langle(\Delta\hat\theta_2)^2\rangle$, from which we obtain the {\em Schwarz inequality}:
\begin{equation}
\left\langle \left(\Delta\hat\theta_1\right)^2\right\rangle \left\langle\left(\Delta\hat\theta_2\right)^2\right\rangle \ge \langle\Delta\hat\theta_1\,\Delta\hat\theta_2\rangle^2
\label{Schwarz}
\end{equation}

Let us treat the simple case of one parameter, and data $x$.  An unbiased estimator $\hat\theta$ of a parameter whose true value is $\theta_0$ is the value $\theta$ which satisfies
\begin{equation}
\langle \theta-\theta_0 \rangle = \int (\theta-\theta_0) L(x;\theta) dx = 0.
\end{equation}
Differentiating with respect to $\theta$ we get
\begin{equation}
\int (\theta-\theta_0) \frac{\partial L}{\partial \theta}dx +\int L(x;\theta) dx= 0.
\end{equation}
The last integral is unity, and we can therefore write
\begin{equation}
\left\langle (\theta-\theta_0) \frac{\partial \ln L}{\partial \theta}\right\rangle = -1,
\end{equation}
and the Schwarz inequality gives
\begin{equation}
\left\langle(\theta-\theta_0)^2 \right\rangle \ge \frac{1}{\left\langle\left(\frac{\partial \ln L}{\partial \theta}\right)^2\right\rangle}
\end{equation}
The final expression is obtained by differentiating $\int L(x;\theta) dx=1$ twice with respect to $\theta$ to show
\begin{equation}
0 = \frac{\partial}{\partial\theta}\int \frac{\partial \ln L}{\partial \theta}L dx =\int \left[\frac{\partial^2 \ln L}{\partial \theta^2}+\left(\frac{\partial \ln L}{\partial \theta}\right)^2 \right] L(x;\theta) dx 
\end{equation}
so we obtain the {\em Cram\' er-Rao inequality}
\begin{equation}
\left\langle(\theta-\theta_0)^2 \right\rangle \ge - \frac{1}{\left\langle\frac{\partial^2 \ln L}{\partial \theta^2}\right\rangle} = \frac{1}{F_{\theta\theta}}.
\end{equation}
Note that for a single variable the conditional error is the same as the marginal error - the Fisher `matrix' has rank 1.
%

%


\subsubsection{Combining experiments}

If the experiments are independent, you can simply add the Fisher matrices (why?).   Note that the marginal error ellipses (marginalising over all but two variables) in the combined dataset can be much smaller than you might expect, given the marginal error ellipses for the individual experiments, because the operations of adding the experimental data and marginalising do not commute.

\subsection{The Gaussian Case}

Let us now explicitly compute the Fisher information matrix for
the case when the probability distribution is Gaussian, {\ie},
where (dropping an irrelevant additive constant $N\ln[2\pi]$)
\begin{equation}\label{GaussianEq}
2\Ell = \ln\det \C + (\x-\vmu)
\C^{-1}(\x-\vmu)^T,
\end{equation}
where in general both the mean vector $\vmu$ and the covariance
matrix
\begin{equation}\label{covdefEq}
\C =\expec{(\x-\vmu)(\x-\vmu)^T}
\end{equation}
depend on the model parameters $\vth$. Although vastly simpler
than the most general situation, the Gaussian case is nonetheless
general enough to be applicable to a wide variety of problems in
cosmology. Defining the data matrix
\begin{equation}\label{DdefEq}
\DD\equiv (\x-\vmu)(\x-\vmu)^T
\end{equation}
and using the matrix identity (see exercises) $\ln\det \C = \Tr\ln
\C$, where Tr indicates trace, we can re-write (\ref{GaussianEq})
as
\begin{equation}\label{GaussianEq2}
2\Ell = \Tr\left[\ln \C + \C^{-1}\DD\right].
\end{equation}
We will use the standard comma notation for derivatives, where for
instance
\begin{equation}\label{CommaDefEq}
\C,_\alpha \equiv {\partial\over\partial\theta_\alpha}\C.
\end{equation}
Since $\C$ is a symmetric matrix for all values of the parameters,
it is easy to see that all the derivatives $\C,_\alpha$, $\C,_{\alpha\beta}$,
will also be symmetric matrices. Using the matrix identities
$(\C^{-1}),_\alpha = -\C^{-1}\C,_\alpha \C^{-1}$ and $(\ln \C),_\alpha = \C^{-1}
\C,_\alpha$ (see exercises), we find
\begin{equation}\label{LiEq}
2\Ell,_\alpha = \Tr\left[\C^{-1} \C,_\alpha -
         \C^{-1}\C,_\alpha \C^{-1}\DD + \C^{-1}\DD,_\alpha\right].
\end{equation}
When evaluating $\C$ and $\vmu$ at the true parameter values,
we have $\expec{\x}=\vmu$ and $\expec{\x\x^T}=\C+\vmu\vmu^T$,
which gives
\begin{equation}\label{DexpecEq}
\cases{ \expec{\DD} &$=\C$,\crr \expec{\DD,_\alpha}   &$=0$,\crr
\expec{\DD,_{\alpha\beta}} &$=\vmu,_\alpha\vmu,_\beta^T + \vmu,_\beta\vmu,_\alpha^T$.\crr }
\end{equation}
Using this and \eq{LiEq}, we obtain
$\expec{\Ell,_\alpha} = 0$. In other words, the ML-estimate is correct
on average in the sense that the average slope of the likelihood
function is zero at the point corresponding to the true parameter
values. Applying the chain rule to \eq{LiEq}, we obtain
\begin{eqnarray}\label{LijEq}
\nonumber 2\Ell,_{\alpha\beta} = \Tr[
&-&\C^{-1}\C,_\alpha\C^{-1}\C,_\beta
 + \C^{-1} \C,_{\alpha\beta}\\
\nonumber
&+& \C^{-1}(\C,_\alpha \C^{-1} \C,_\beta + \C,_\beta \C^{-1} \C,_\alpha)\C^{-1} \DD\\
\nonumber
&-& \C^{-1}(\C,_\alpha \C^{-1} \DD_{,\beta}+\C,_\beta \C^{-1} \DD_{,\alpha})\\
&-& \C^{-1}\C,_{\alpha\beta} \C^{-1} \DD
 + \C^{-1} \DD,_{\alpha\beta}
].
\end{eqnarray}
Substituting this and \eq{DexpecEq} into
\eq{FisherDefEq} and using the trace identity $\Tr [\A\B] = \Tr
[\B\A]$, many terms drop out and the Fisher information matrix
reduces to simply
\begin{equation}\label{FullFisher}
\F_{\alpha\beta} = \expec{\Ell,_{\alpha\beta}} = {1\over 2}\Tr[\C^{-1}\C_{,\alpha}
\C^{-1}\C_{,\beta} + \C^{-1}\M_{\alpha\beta}],
\end{equation}
where we have defined the matrix $\M_{\alpha\beta}\equiv \expec{\DD,_{\alpha\beta}}=
\vmu,_\alpha\vmu,_\beta^T + \vmu,_\beta\vmu,_\alpha^T$.

The Fisher matrix requires no data.

This result is extremely powerful.  If the data have a (multivariate) gaussian distribution
(and the errors can be correlated; $\C$ need not be diagonal), and you know
how the means $\vmu$ and the covariance matrix $\C$ depend on the parameters, you can
calculate the Fisher Matrix {\em before you do the experiment}.  The Fisher Matrix
gives you the expected errors, so you know how well you can expect to do if you do a particular
experiment, and you can then design an experiment to give you, for example, the best (marginal) error on
the parameter you are most interested in.  

Note that if the prior is not uniform, then you can simply add a
`prior matrix' to the Fisher matrix before inversion.  Fig.
\ref{VISTA} shows an example, where a prior from CMB experimental
results has been added to a hypothetical 3D weak lensing survey.

Treat the Fisher errors as a one-way test:  you might not achieve errors which are this small, but you won't do better. So if you want to measure some quantity with an accuracy of a metre, and a Fisher analysis tells you the error bar is the size of Belgium, give up.

Finally, note that this analysis assumes that the data do not depend on the parameters.  Normally this is the case - you simply measure the data, right?  Be careful - if you construct a new dataset (e.g. estimates of a galaxy correlation function), you may need to assume what the parameters are (e.g. what the cosmology is, to calculate the pair separations).  In this case, you should assume a fiducial set of cosmological parameters to calculate your dataset, and not change the data as you change the parameters.  You then need to do more work on the theory side, but the error bars should be robust.

\subsubsection{Reparametrisation}

Finally we mention that sometimes the gaussian approximation for the likelihood surface is not a very good approximation.  With a good theoretical model, it is possible to make nonlinear transformations of the parameters to make the likelihood more gaussian.  See \cite{Kosowsky02} for more details.

\subsubsection{iCosmo: a great resource}

icosmo.org has a web-based calculator for Fisher matrices for cosmology.  You can download results, or even the source code, to compute Fisher matrices for various experiments in lensing, BAOs, supernovae etc, or custom-design your own survey.   It also gives many other useful things, such as lensing power spectra, angular diameter distances etc.

\section{Numerical methods}

If the problem has only two or three parameters, then it may be possible to evaluate the likelihood on a sufficiently fine grid to be able to locate the peak and estimate the errors.   If the dimensionality of the parameter space is very large, then, as the number of grid points grows exponentially with dimension, it becomes rapidly unfeasible to do it this way.    In fact, it's very inefficient to do this anyway, as typically most of the hypervolume has very small likelihood so is of little interest.  There are various ways to sample the likelihood surface more efficiently,  concentrating the points more densely where the likelihood is high.  We cover here the most common method (MCMC), and a relatively new method (to cosmology), Hamiltonian Monte Carlo, which could take over, as it seems more efficient in cases studied.

\subsection{Monte Carlo Markov Chain (MCMC) method}

The aim of MCMC is to generate a set of points in the parameter space whose distribution function is the same as the {\em target density}, in this case the likelihood, or more generally the posterior.   MCMC makes random drawings, by moving in parameter space in a Markov process - i.e. the next sample depends on the present one, but not on previous ones.  By design, the resulting Markov Chain of points samples the posterior, such that the density of points is proportional to the target density (at least asymptotically), so we can estimate all the usual quantities of interest from it (mean, variance, etc).  The number of points required to get good estimates is said to scale linearly with the number of parameters, so very quickly becomes much faster than grids as the dimensionality increases.   In cosmology, we are often dealing with around 10-20 parameters, so MCMC has been found to be a very effective tool.

The  target density is approximated by a set of delta functions (you may need to normalise)
\be
p(\vth) \simeq \frac{1}{N}\sum_{i=1}^N \delta(\vth-\vth_i) 
\ee
from which we can estimate any integrals (such as the mean, variance etc.):
\be
\left\langle f(\vth)\right\rangle \simeq \frac{1}{N}\sum_{i=1}^N f(\vth_i).
\ee

The basic procedure to make the chain is to generate a new point $\vth^*$ from the present point $\vth$ (by taking some sort of step),  and accepting it as a new point in the chain with a probability which depends on the ratio of the new and old target densities.    The distribution of steps is called the {\em proposal distribution}.  The most popular algorithm is the {\em Metropolis-Hastings} algorithm, where the probability of acceptance is
\be
p({\rm acceptance}) = min\left[1,\frac{p(\vth^*)q(\vth^*|\vth)}{p(\vth)q(\vth|\vth^*)}\right]
\ee
where the proposal distribution function is $q(\vth*|\vth)$ for a move from $\vth$ to $\vth^*$.

If the proposal distribution is symmetric (as is often the case), the algorithm simplifies to the {\em Metropolis algorithm}:
\be
p({\rm acceptance}) = min\left[1,\frac{p(\vth^*)}{p(\vth)}\right].
\ee

\begin{itemize}
\item{Choose a random initial starting point in parameter space, and compute the target density.}
\item{Repeat:}
\item{ Generate a step in parameter space from a {\em proposal distribution}, generating a new trial point for the chain.}
\item{\indent Compute the target density at the new point, and accept it (or not) with the Metropolis-Hastings algorithm.}
\item{\indent If the point is not accepted, {\em the previous point is repeated in the chain}\footnote{It is a common mistake  to neglect to do this}.}
\item{End Repeat:}
\end{itemize}
The easy bits:  this is trivial to code - you might just take a top-hat proposal distribution in each parameter direction, and it should work.   The harder parts are (even in the tophat case): choosing an efficient proposal distribution; dealing with {\em burn-in} and {\em convergence}.

\subsubsection{Proposal distribution}

If the proposal distribution is small, in the sense that the typical jump is small, then the chain may take a very long time to explore the target distribution, and it will be very inefficient.  Since the target density hardly changes, almost all points are accepted, but it still takes forever.   This is an example of poor {\em mixing}.  If the proposal distribution is too large, on the other hand, then the parameter space is explored, but the trial points are often a long way from the peak, at places where the target density is low.  This is very inefficient as well, since they are almost always rejected by the Metropolis-Hastings algorithm.  So what is best?  You might expect the chain to do well if the proposal distribution is `about the same size as the peak in the target density', and you would be right.  In fact, a very good option is to draw from a multivariate gaussian with the Fisher matrix as Hessian.   However, having said that, if you want something quick to code (but sub-optimal), a top-hat of about the right dimensions will do a decent job.  You might have to run a preliminary chain or two to get an idea of the size of the target, but the rules say you are not allowed to change the proposal distribution in a chain, so once you have decided you must throw away your chains and begin again.

\subsubsection{Burn-in and convergence}

\begin{figure}
\centering
\includegraphics[height=11cm, angle=90]{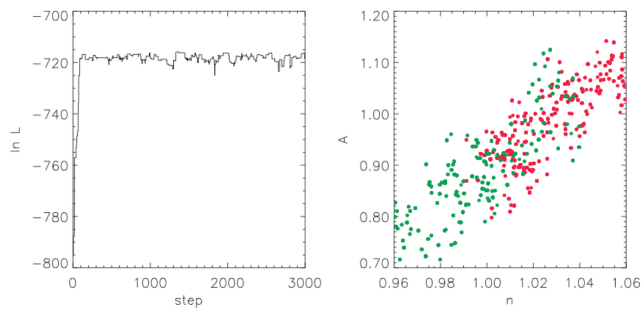}
\caption{Examples of unconverged chains.  The left panel suggests the chain has converged, but the right panel shows the chain and a second one, also apparently converged, but showing clearly different distributions.  From Verde et al. ApJS, 148, 195  (2003).}
\label{MCMCunconverged}       
\end{figure}

Theory indicates that the chain should fairly sample the target distribution once it has converged to a stationary distribution.  This means that the early part of the chain (the `burn-in' are ignored, and the dependence on the starting point is lost.  {\em It is vitally important to have a convergence test}. Be warned that the points in a MCMC chain are correlated, and the chain can appear to have converged when it has not (one can reduce this problem by evaluating the correlation function of the points, and `thinning' them by (say) taking only every third point (or whatever is suggested by the correlation analysis).  Fig. \ref{MCMCunconverged} shows one such example.   The classic test is the {\em Gelman-Rubin} (1992) convergence criterion.  Start $M$ chains, each with $2N$ points, starting at well-separated parts of parameter space.  In this example, the first $N$ are discarded as burn-in.  

The idea is that you have two ways to estimate the mean of the parameters - either treat the combined chains as a single dataset, or look at the means of each chain.  If the chains have converged, these should agree within some tolerance.

Following \cite{VerdeNotes}, let $\theta_I^J$ represent the point in parameter space in position $I$ of chain $J$.  Compute the mean of each chain ($J$):
\be
\bar\theta^J \equiv \frac{1}{N}\sum_{I=1}^{N}\theta_I^J
\ee
and the mean of all the chains
 \be
\bar\theta \equiv \frac{1}{NM}\sum_{I=1}^N \sum_{J=1}^{M}\theta_I^J.
\ee
The chain-to-chain variance $B$ is
\be
B = \frac{1}{(M-1)}\sum_{J=1}^{M}(\bar \theta^J-\bar\theta)^2
\ee
and the average variance of each chain is
\be
W = \frac{1}{M(N-1)}\sum_{I=1}^{N}\sum_{J=1}^M(\theta_I^J-\bar\theta^J)^2.
\ee
Under convergence, $W$ and $B/N$ should agree.

The weighted estimate of the variance,
\be
\sigma^2 = \frac{N-1}{N}W + \frac{B}{N}
\ee
overestimates the true variance if the starting distribution is overdispersed.

Accounting for the variance of the means gives an estimator of the variance
\be 
V = \sigma^2 + \frac{B}{NM}
\ee

$V$ is an overestimate, and $W$ is an underestimate.  The ratio of the two estimates is
\be
\hat R = \frac{\left(\frac{N-1}{N}\right) + B\left(1+\frac{1}{M}\right)}{W}.
\ee
$R$ should approach unity as convergence is achieved.  How close to $R=1$? Opinions differ; I have seen suggestions to run the chain until the values of $\hat R$ are always $<1.03$, or $<1.2$, but a proof would be nice.

\subsubsection{CosmoMC}

Excellent resource.  MCMC sampler with cosmological data (CMB + support for LSS, SNe). http://cosmologist.info/cosmomc/

For more details on MCMC in general, see \cite{Gilks,Lewis,Verde}.

\subsection{Hamiltonian Monte Carlo}

As we've seen, the proposal distribution has to be reasonably finely tuned to ensure good mixing, but not be too inefficient.  What we would really like is to be able to take rather big jumps, so the chain is well-mixed, but in such a way that the probability of acceptance of each point is still high.  Hamiltonian (or Hybrid) Monte Carlo (HMC) tries to do this, via a rather clever trick.  In practice, it seems that it can be about $M$ times faster than MCMC in $M$ dimensions.  Typical applications report that 4 times shorter chains give the same accuracy as MCMC.  Originally developed for particle physics \cite{Duan}, there is a nice exposition in astrophysics by Hajian \cite{Hajian06}.

HMC works by sampling from a {\em larger} parameter space than we want to explore, by introducing $M$ {\em auxiliary variables}, one for each parameter in the model.  To see how this works, imagine each of the parameters in the problem as a coordinate.  HMC regards the target distribution which we seek as an effective potential in this coordinate system, and for each coordinate it generates a generalised momentum.  i.e. it expands the parameter space from its original `coordinate space' to a phase space, in which there is a so-called `extended' target distribution. It tries to sample from the extended target distribution in the $2M$ dimensions.  It explores this space by treating the problem as a dynamical system, and evolving the phase space coordinates by solving the dynamical equations.  Finally, it ignores the momenta (marginalising, as in MCMC), and this gives a sample of the original target distribution.  

There are several advantages to this approach.  One is that the probability of acceptance of a new point is high - close to unity; secondly, the system can make big jumps, so the mixing is better and the convergence faster.  In doing the big jumps, it does some additional calculations, but these do not involve computing the likelihood, which is typically computationally expensive.

Let us see how it works.  If the target density in $M$ dimensions is $p(\vth)$, then we define a potential
\be
U(\vth) \equiv -\ln p(\vth).
\ee
For each coordinate $\theta_\alpha$, we generate a momentum $u_\alpha$, conveniently from a normal distribution with zero mean and unit variance, so the $M$-dimensional momentum distribution is a simple multivariate gaussian which we denote ${\cal N}(\bu)$.  We define the kinetic energy
\be
K(\bu) \equiv \frac{1}{2}\bu^T\bu,
\ee
and the Hamilitonian is 
\be
H(\vth,\bu) \equiv U(\vth) + K(\bu).
\ee
The trick is that we generate chains to sample the {\em extended target density} 
\be
p(\vth,\bu)  = \exp\left[-H(\vth,\bu)\right].
\ee
Since this is separable,
\be
p(\vth,\bu) = \exp\left[-U(\vth)\right]\exp\left[-K(\bu)\right] \propto p(\vth){\cal N}(\bu)
\ee
and if we then marginalise over $\bu$ by simply ignoring the $\bu$ coordinates attached to each point in the chain (just as in MCMC), the resulting marginal distribution samples the desired target distribution $p(\vth)$.  This is really very neat.

The key is that if we {\em exactly} solve the Hamiltonian equations
\begin{eqnarray}
\dot\theta_\alpha &=& u_\alpha \nonumber\\
\dot u_\alpha &=& -\frac{\partial H}{\partial \theta_\alpha}
\end{eqnarray}
then $H$ remains invariant, so the extended target density is always the same, and the acceptance is unity.  Furthermore, we can integrate the equations for a long time if we wish, decorrelating the points in the chain.  

There are several issues to consider.  
\begin{itemize}
\item{We seem to need the target density to define the potential, but this is what we are looking for.  We need to approximate it.}
\item{The aim is to do this fast, so we do not want to do many operations before generating a new sample.  An easy way to achieve this is to employ a simple integrator (e.g. leap-frog; even Euler's method might do) and take several fairly big steps before generating a new point in the chain. }
\item{We need to ensure we explore the extended space carefully.}
\end{itemize}

The result of the two approximations is that $H$ will not be quite constant.  We deal with this as in MCMC by using the Metropolis algorithm, accepting the new point $(\vth^*,\bu^*)$ with a probability
\be
min\left\{1,\exp\left[-H(\vth^*,\bu^*)+H(\vth,\bu)\right]\right\},
\ee
otherwise we repeat the old point $(\vth,\bu)$ as usual.

How do we approximate?  We want the gradients to be cheap to compute.  It is usual to generate an approximate analytical potential by running a relatively short MCMC chain, computing the covariance of the points in the chain, and approximating the distribution by a multivariate gaussian by a multivariate gaussian with the same covariance.  The gradients are then consequently easy to compute analytically.

\begin{figure}
\centering
  \includegraphics[width=0.45\textwidth]{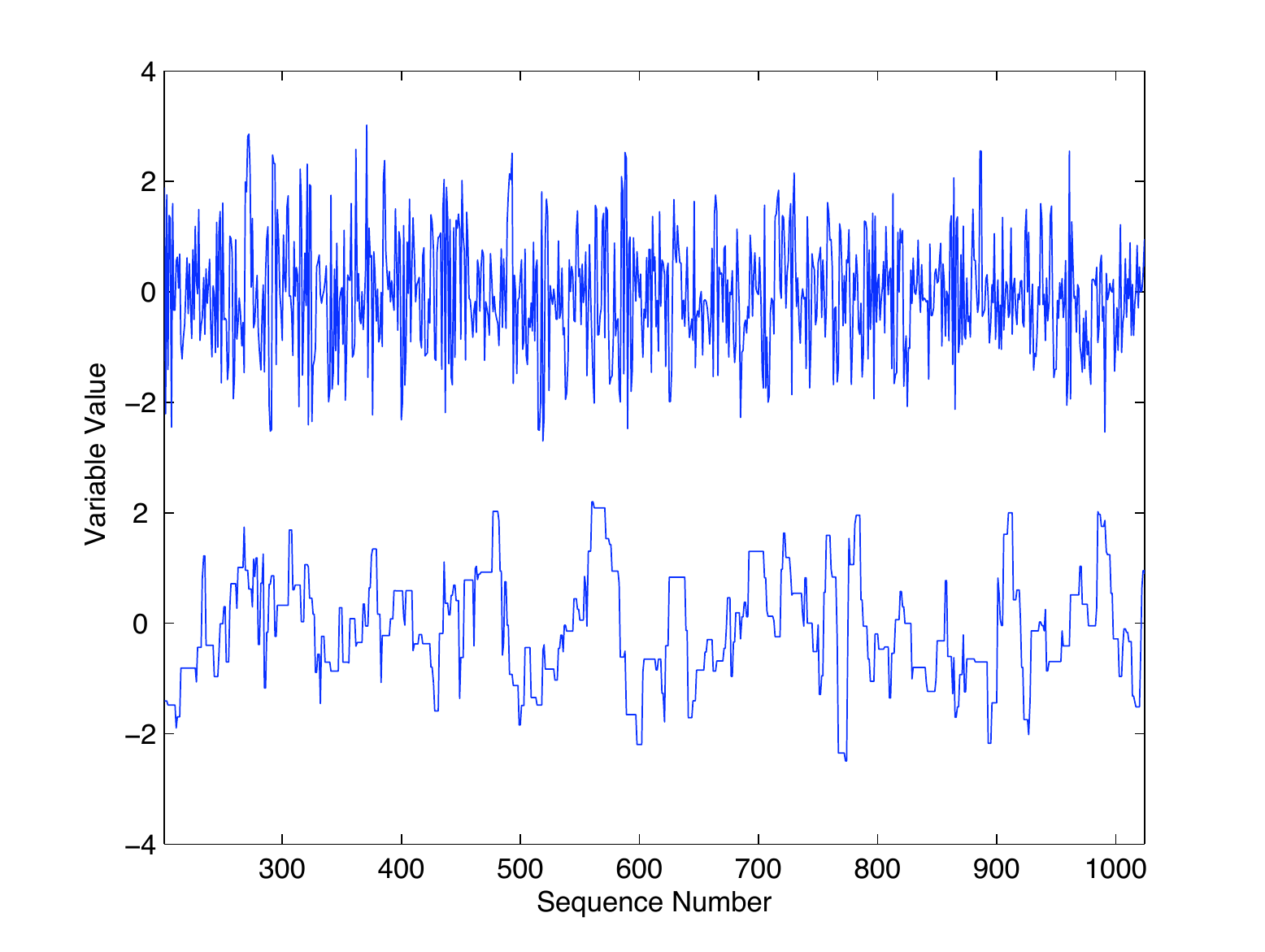}\\
  \caption{A comparison of HMC sampling (top) and MCMC sampling (bottom).  Note that the computer time required to generate each point in the HMC sampling will be larger than that of MCMC, so the actual gains are less than appears. From \cite{Hajian06}.}
  \label{HMC}
\end{figure}

The last point is that if we change the momentum only with Hamilton's equations of motion, we will restrict ourselves to a locus in phase space, and the target distribution will not be properly explored.  To avoid this, a new momentum is generated randomly when each point in the chain is generated.   The art is to choose a good step in the integration, and the number of steps to take before generating a new point.  Perhaps unsurprisingly, choosing these such that the new point differs from the previous one by about the size of the target peak works well. Thinning can be performed, and a convergence test must still be applied.  Fig. \ref{HMC} shows a comparison between HMC and MCMC for a simple case of a 6D gaussian target distribution, from \cite{Hajian06}.

\section{Model Selection}

Model selection is in a sense a higher-level question than parameter estimation.  In parameter estimation, one assumes a theoretical model within which one interprets the data, whereas in model selection, one wants to know which theoretical framework is preferred, given the data (regardless of the parameter values).  The models may be completely different (e.g. compare Big Bang with Steady State, to use an old example), or variants of the same idea.  E.g. comparing a simple cosmological model where the Universe is assumed flat and the perturbations are strictly scale-invariant ($n=1$), with a more general model where curvature is allowed to vary and the spectrum is allowed to deviate from scale-invariance.  The sort of question asked here is essentially `Do the data require a more complex model?'.  Clearly in the latter type of comparison $\chi^2$ itself will be of no use - it will always reduce if we allow more freedom.  There are frequentist ways to try and answer these questions, but we are all by now confirmed Bayesians\footnote{If not, please leave the room}, so will approach it this way.

\subsection{Bayesian evidence}

The Bayesian method to select between models (e.g. General Relativity, \& modified gravity) is to consider the
Bayesian evidence ratio.  Essentially we want to know if, given the data, there is evidence that we need to expand the space of gravity models beyond GR.  Assuming non-commital priors for the models (i.e. the same a priori probability), the probability of the models given the data is simply proportional to the evidence.

We denote two competing models by $M$ and $M'$.  We assume that $M'$
is a simpler model, which has fewer ($n'<n$) parameters in it.  We
further assume that it is {\em nested} in Model $M'$, i.e. the $n'$
parameters of model $M'$ are common to $M$, which has $p\equiv n-n'$
extra parameters in it. These parameters are fixed to fiducial values in $M'$.

We denote by $\bx$ the data vector, and by $\vth$ and $\vth'$ the
parameter vectors (of length $n$ and $n'$).

Apply Rule 1: Write down what you want to know.  Here it is $p(M|\bx)$ - the probability of the model, given the data.

The posterior probability of each model comes from Bayes' theorem:
\begin{equation}
p(M|\bx) = \frac{p(\bx|M)p(M)}{p(\bx)}
\end{equation}
and similarly for $M'$.  By marginalisation $p(\bx|M)$, known as the
{\em Evidence}, is
\begin{equation}
p(\bx|M) = \int d\vth\,p(\bx|\vth M)p(\vth|M),
\end{equation}
which should be interpreted as a multidimensional integration. Hence
the posterior relative probabilities of the two models, regardless
of what their parameters are\footnote{If a model has no parameters, then the integral is simply replaced by $p(\bx|M)$}, is
\begin{equation}
\frac{p(M'|\bx)}{p(M|\bx)}=\frac{p(M')}{p(M)}\frac{\int
d\vth'\,p(\bx|\vth' M')p(\vth'|M')}{\int
d\vth\,p(\bx|\vth M)p(\vth|M)}.
\end{equation}
With non-committal priors on the models, $p(M')=p(M)$, this ratio
simplifies to the ratio of evidences, called the {\em Bayes Factor},
\begin{equation}
B \equiv \frac{\int d\vth'\,p(\bx|\vth' M')p(\vth'|M')}{\int
d\vth\,p(\bx|\vth M)p(\vth|M)}.
\end{equation}
Note that the a complicated model $M$ will (if $M'$ is nested) inevitably lead to a
higher likelihood (or at least as high), but the evidence will
favour the simpler model if the fit is nearly as good, through the
smaller prior volume.

We assume uniform (and hence separable) priors in each parameter,
over ranges $\Delta\vth$ (or $\Delta\vth'$).  Hence
$p(\vth|M)=(\Delta\vth_1\ldots \Delta\vth_n)^{-1}$ and
\begin{equation}
B = \frac{\int d\vth'\,p(\bx|\vth',M')}{\int
d\vth\,p(\bx|\vth,M)}\,\frac{\Delta\vth_1\ldots\Delta\vth_n}{\Delta\vth'_1\ldots\Delta\vth'_{n'}}.
\label{Bnew}
\end{equation}
Note that if the prior ranges are not large enough to contain
essentially all the likelihood, then the position of the boundaries
would influence the Bayes factor.  In what follows, we will assume
the prior range is large enough to encompass all the likelihood.

In the nested case, the ratio of prior hypervolumes simplifies to
\begin{equation}
\frac{\Delta\vth_1\ldots\Delta\vth_n}{\Delta\vth'_1\ldots\Delta\vth'_{n'}}=\Delta\vth_{n'+1}\ldots
\Delta\vth_{n'+p},
\end{equation}
where $p\equiv n-n'$ is the number of extra parameters in the more
complicated model.

Here we see the problem.  The evidence requires a multidimensional integration over the likelihood and prior, and this may be {\em very} expensive to compute.  There are various ways to simplify this.  One is analytic - follow the Fisher approach and assume the likelihood is a multivariate gaussian, others are numerical, such as nested sampling, where one tries to sample the likelihood in an efficient way.  There are others, but we will focus on these.  Note that shortcuts with names such as AIC and BIC may be unreliable as they are based on the best-fit $\chi^2$, and from a Bayesian perspective we want to know how much parameter space would give the data with high probability.  See \cite{Liddle07} for more discussion.

\subsection{Laplace approximation}

The Bayes factor in equation (\ref{Bnew}) still depends on the specific dataset $\bx$.  For future experiments, we do not yet have the data, so we compute the expectation value of the Bayes factor, given the statistical properties of $\bx$.  The expectation is computed over the distribution of $\bx$ for the correct model (assumed here to be $M$).  To do this, we make two further approximations: first we note that $B$ is a ratio, and we approximate $\langle B\rangle$ by the ratio of the expected values, rather than the expectation value of the ratio.  This should be a good approximation if the evidences are sharply peaked.

We also make the Laplace approximation, that the expected likelihoods are given by multivariate Gaussians.  For example,
\begin{equation}
\langle p(\bx|\vth,M)\rangle = L_0 \exp\left[-\frac{1}{2}(\vth-\vth_0)_\alpha
\F_{\alpha\beta}(\vth-\vth_0)_\beta\right]
\end{equation}
and similarly for $\langle p(\bx|\vth',M')\rangle$.  This assumes that a Taylor
expansion of the likelihood around the peak value to second order
can be extended throughout the parameter space. $F_{\alpha\beta}$ is
the Fisher matrix, given for Gaussian-distributed data by equation ({\ref{FullFisher}):
\begin{equation}
\F_{\alpha\beta}=\frac{1}{2}{\rm
Tr}\left[\C^{-1}\C_{,\alpha}\C^{-1}\C_{,\beta}+\C^{-1}(\mu_{,\beta}\mu^T_{,\alpha}+\mu_{,\alpha}\mu^T_{,\beta})\right].
\end{equation}
$C$ is the covariance matrix of the data, and $\mu$ its mean (no
noise). Commas indicate partial derivatives with respect to the parameters.
For the correct model $M$, the peak of the expected likelihood is located at the true parameters $\vth_0$.  Note, however, that for the incorrect model $M'$, the peak of the expected likelihood is not in general at the true parameters (see Fig. \ref{offsetfig} for an illustration of this).  This arises because the likelihood in the numerator of equation (\ref{Bnew}) is the probability of the dataset $\bx$ given incorrect model assumptions.

The Laplace approximation is routinely used in forecasting marginal errors in parameters, using the Fisher matrix.  Clearly the approximation may break down in some cases, but for Planck, the Fisher matrix errors are reasonably close to (within 30\% of) those computed with Monte Carlo Markov Chains.

If we assume that the posterior probability densities are small at
the boundaries of the prior volume, then we can extend the
integrations to infinity, and the integration over the multivariate
Gaussians can be easily done.  This gives, for $M$, $(2\pi)^{n/2}
(\det{\F})^{-1/2}$, so for nested models,
\begin{equation}
\langle B \rangle=
(2\pi)^{-p/2}\frac{\sqrt{\det{\F}}}{\sqrt{\det{\F'}}}\frac{L'_0}{L_0}\Delta\vth_{n'+1}\ldots
\Delta\vth_{n'+p}.
\end{equation}
An equivalent expression was obtained, using again the Laplace
approximation by \cite{Lazarides}. The point here is that with the
Laplace approximation, one can compute the $L'_0/L_0$ ratio from the
Fisher matrix.  To compute this ratio of likelihoods, we need to
take into account the fact that, if the true underlying model is
$M$, in $M'$ (the incorrect model), the maximum of the expected
likelihood will not in general be at the correct values of the
parameters (see Fig. \ref{offsetfig}). The $n'$ parameters shift
from their true values to compensate for the fact that, effectively,
the $p$ additional parameters are being kept fixed at incorrect
fiducial values. If in $M'$, the additional $p$ parameters are
assumed to be fixed at fiducial values which differ by
$\delta\psi_\alpha$ from their true values, the others are shifted
on average by an amount which is readily computed under the
assumption of the multivariate Gaussian likelihood:
\begin{equation}
\delta\vth'_\alpha =
-(\F'^{-1})_{\alpha\beta}\G_{\beta\zeta}\delta\psi_\zeta \qquad
\alpha,\beta=1\ldots n', \zeta=1\ldots p \label{offset}
\end{equation} where
\begin{equation}
\G_{\beta\zeta}=\frac{1}{2}{\rm
Tr}\left[\C^{-1}\C_{,\beta}\C^{-1}\C_{,\zeta}+\C^{-1}(\mu_{,\zeta}\mu^T_{,\beta}+\mu_{,\beta}\mu^T_{,\zeta})\right],
\end{equation}
which we recognise as a subset of the Fisher matrix. For clarity, we
have given the additional parameters the symbol $\psi_\zeta; \
\zeta=1 \ldots p$ to distinguish them from the parameters in $M'$.

\begin{figure}
\centering
  \includegraphics[width=0.45\textwidth]{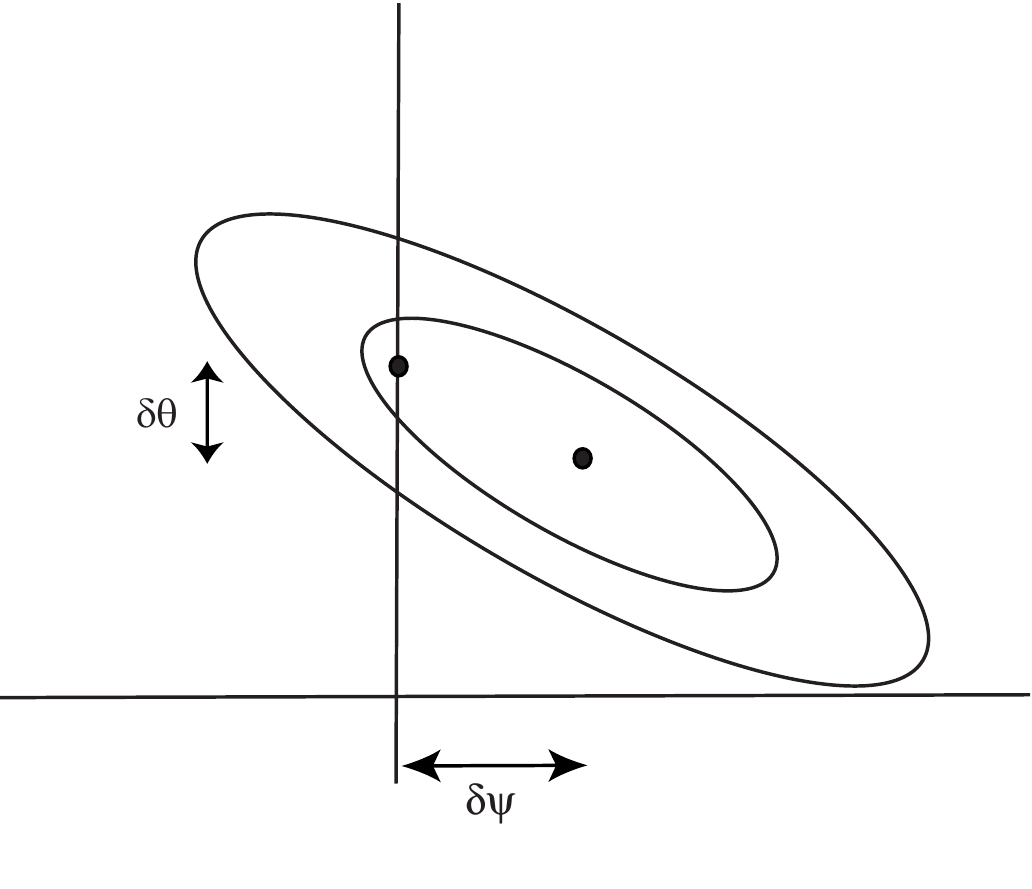}\\
  \caption{Illustrating how assumption of a wrong parameter value can influence the
  best-fitting value of other model parameters.  Ellipses represent iso-likelihood surfaces,
  and here in the simpler model, the parameter on the horizontal axis is assumed to take the
  value given by the vertical line. Filled circles show the true parameters in the more complicated model,
  and the best-fit parameters in the simpler model. From \cite{Heavens07}.}
  \label{offsetfig}
\end{figure}
With these offsets in the maximum likelihood parameters in model
$M'$, the ratio of likelihoods is given by
\begin{equation}
L'_0 = L_0 \exp\left(-\frac{1}{2}\delta\vth_\alpha \F_{\alpha\beta}
\delta\vth_\beta\right)
\end{equation}
where the offsets are given by $\delta\vth_\alpha =
\delta\vth'_\alpha$ for $\alpha\le n'$ (equation \ref{offset}),
and $\delta\vth_\alpha = \delta\psi_{\alpha-n'}$ for $\alpha>n'$.

The final expression for the expected Bayes factor is then
\begin{equation}
\langle B \rangle =
(2\pi)^{-p/2}\frac{\sqrt{\det{\F}}}{\sqrt{\det{\F'}}}\exp\left(-\frac{1}{2}\delta\vth_\alpha
\F_{\alpha\beta}\delta\vth_\beta\right)\prod_{q=1}^p\Delta\vth_{n'+q}.
\label{Final}
\end{equation}
Note that $\F$ and $\F^{-1}$ are $n \times n$ matrices, $\F'$ is $n'
\times n'$, and $\G$ is an $n' \times p$ block of the full $n \times
n$ Fisher matrix $\F$.  The expression we find is a specific example
of the Savage-Dickey ratio (see e.g. \cite{Trotta07}).  For a nested model with a single additional parameter $\vth_i$,
\be
\langle B \rangle = \frac{p(\vth_i|\bx)}{p(\vth_i)}.
\ee

Here we explicitly use
the Laplace approximation to compute the offsets in the parameter
estimates which accompany the wrong choice of model, and compute the evidence ratio explicitly.  Finally, note that this is the expected 
evidence ratio (nearly); it does not address the issue of what the distribution of evidence ratios should be.

Note that the `Occam's razor' term,
common in evidence calculations, is to some extent encapsulated in the term
$(2\pi)^{-p/2}\frac{\sqrt{\det{\F}}}{\sqrt{\det{\F'}}}$,
multiplied by the prior product: models with more parameters are
penalised in favour of simpler models, unless the data demand
otherwise. In cases where the Laplace
approximation is not a good one, other techniques must be used, at
more computational expense.  

It is perhaps worth remarking that Occam's razor appears not to be fully incorporated into this term, as can be seen by considering a situation where the data do not depend at all on an additional parameter.  In this case, the Bayesian evidence ratio is unity, so disappointingly no preference is shown at all. 

As an example, consider testing General Relativity against other gravity theories, which predict a different growth rate of perturbations, $d\ln\delta/d\ln\Omega_m = \gamma$, where $\gamma=0.55$ for GR, and (for example) $\gamma=0.68$ for 
a flat DGP braneworld model.  This can be probed with weak lensing,  and we ask the question do the data favour a model where $\gamma$ is a free parameter, rather than being fixed at $0.55$?  

We take a prior range $\Delta\gamma=1$, and we ask the question of how different the growth
rate of a modified-gravity model would have to be for these
experiments to be expected to favour a relaxation of the gravity  model from General
Relativity.  This is shown in Fig.\ref{dgamma}.  It shows how
the expected evidence ratio changes with progressively greater
differences from the General Relativistic growth rate.  We see that
a next-generation weak lensing survey could even distinguish `strongly'
$\delta\gamma=0.048$.  Note that changing the prior range
$\Delta\gamma$ by a factor 10 changes the numbers
by $\sim 0.012$, so the dependence on the prior range is rather
small.

If one prefers to ask a frequentist question, then a combination of
WL+{\em Planck}+BAO+SN should be able to distinguish
$\delta\gamma=0.13$, at $10.6\sigma$.   Alternatively, one can calculate the expected error on $\gamma$ \cite{Amendola} within the extended model $M$.  In this section, we are asking a slightly different question of whether the data demand that a wider class of models needs to be considered at all, rather than estimating a parameter within that wider class.

\begin{figure}
\centering
  \includegraphics[width=0.45\textwidth]{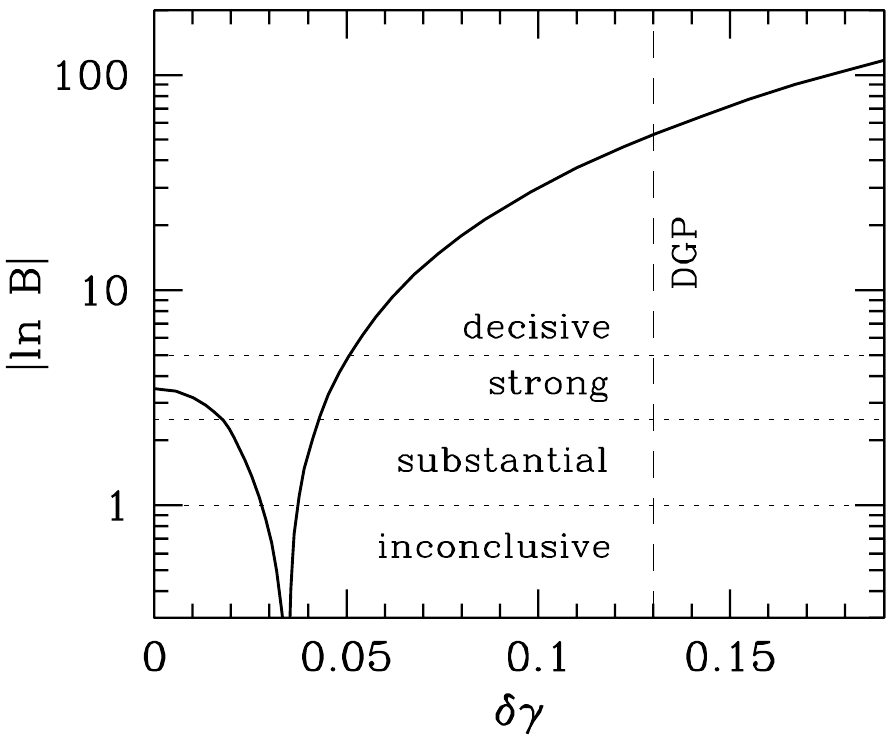}\\
  \caption{The expected value of $|\ln \langle B\rangle|$ from a future large-scale deep weak lensing survey, as might be done with Euclid, JDEM or LSST, 
  in combination with CMB
constraints from \emph{Planck}, as a function of the difference in
the growth rate between the modified-gravity model and General
Relativity.  The crossover at small $\delta\gamma$ occurs because
Occam's razor will favour the simpler (General Relativity) model
unless the data demand otherwise. To the left of the cusp, GR would
be likely to be preferred by the data. The dotted vertical line
shows the offset of the growth factor for the flat DGP model. The descriptors are the
terminology of Jeffreys (1961) \cite{Jeffreys61}.  From \cite{Heavens07}}
  \label{dgamma}
\end{figure}

The case for a large, space-based 3D weak lensing survey is
strengthened, as it offers the possibility of conclusively
distinguishing Dark Energy from at least some modified gravity
models.


\subsection{Numerical sampling methods}

In order to compute the evidence numerically, the prior volume needs to be sampled, in much the same way as in 
parameter estimation, except the requirements are slightly more stringent.  MCMC could be used, although there are claims
that it is not good at exploring the parameter space adequately.  I find these claims puzzling, as in evidence calculations we are doing an $N$-dimensional integral,
which is not so different from doing the $(N-1)$-dimensional integral in MCMC to get marginal errors.   However, here are a couple of other sampling techniques.

\subsubsection{The VEGAS algorithm, with rotation}

This is suitable for single-peaked likelihoods.  It is in Numerical Recipes, but needs a modification for efficiency. Essentially, one seeks to sample from a distribution which is close to the target distribution (sampling from a different distribution is called  {\em importance sampling}), but one does not know what it is yet.   One can do this iteratively, sampling from the prior first, then estimating the posterior to get a first guess at the posterior, and using that to refine the sampling distribution. 

Now, one does not want to draw randomly from a {\em non-separable} function of the parameters, as a moment's thought will tell you that this is computationally very expensive, so one seeks a separable function, so one can then draw the individual parameters one after the other from $N$ distributions.  This works well if the target distribution is indeed separable in the parameters, but not otherwise.

The key extra step \cite{Serra07} is to rotate the parameter axes, which can be done by computing (for example) the moments of the distribution (essentially the moment-of-inertia) after any step, and diagonalising it to find the eigenvectors.

The probability to be sampled is
\begin{equation}
p(\vec\theta) \propto g_1(\theta_1)g_2(\theta_2)\ldots
g_M(\theta_M).
\end{equation}
where it can be shown that
\begin{equation}
g_\alpha(\theta_\alpha) \propto \sqrt{\int_{\beta\ne \alpha} d^{M-1}\theta_\beta
\frac{f^2(\vec\theta)}{\prod_{\beta\ne \alpha} g_\beta(\theta_\beta)}}
\end{equation}
where $f$ is the desired target distribution.  Note that $g$ depends on all other $g$s.  $g$ can be improved iteratively.

\begin{figure}
\centering
\includegraphics[height=8cm]{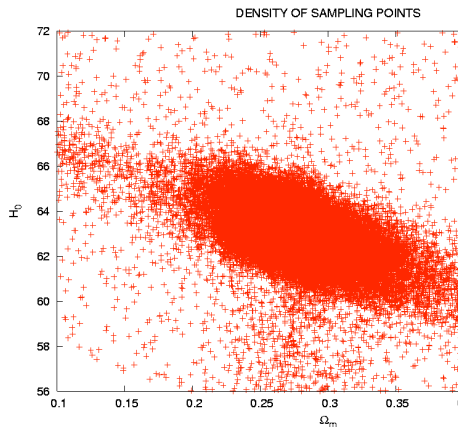}
\caption{The VEGAS Sampling algorithm, applied to supernova data (from \cite{Serra07}}
\label{VEGAS}     
\end{figure}

\subsubsection{Nested sampling}

Other sampling methods can also be very effective (see e.g. \cite{Trotta07, Beltran05,Hobson,MPL06}).  Nested sampling was introduced by Skilling \cite{Skilling}.  One samples from the prior volume, and gradually concentrates more points near the peak of the likelihood distribution, by repeatedly replacing the point with the lowest target density by one drawn strictly from the prior volume with higher target density.  This has proved effective for cosmological model selection.  I will not go into details here, except to say that the key is in drawing the new point from a suitable subset of the prior volume.  This must be increasingly smaller as the points get more confined, otherwise the trial points will be rejected with increasing frequency and it becomes very inefficient, but it must also be a large enough subset that the entire prior volume above the lowest target density is almost certainly included inside.  For further details, see the original papers.  Note that, for multimodal target distributions, a modification called MultiNest exists \cite{Feroz}.   CosmoNest and MultiNest are publicly-available additions to CosmoMC.

\begin{figure}
\centering
\includegraphics[height=8cm]{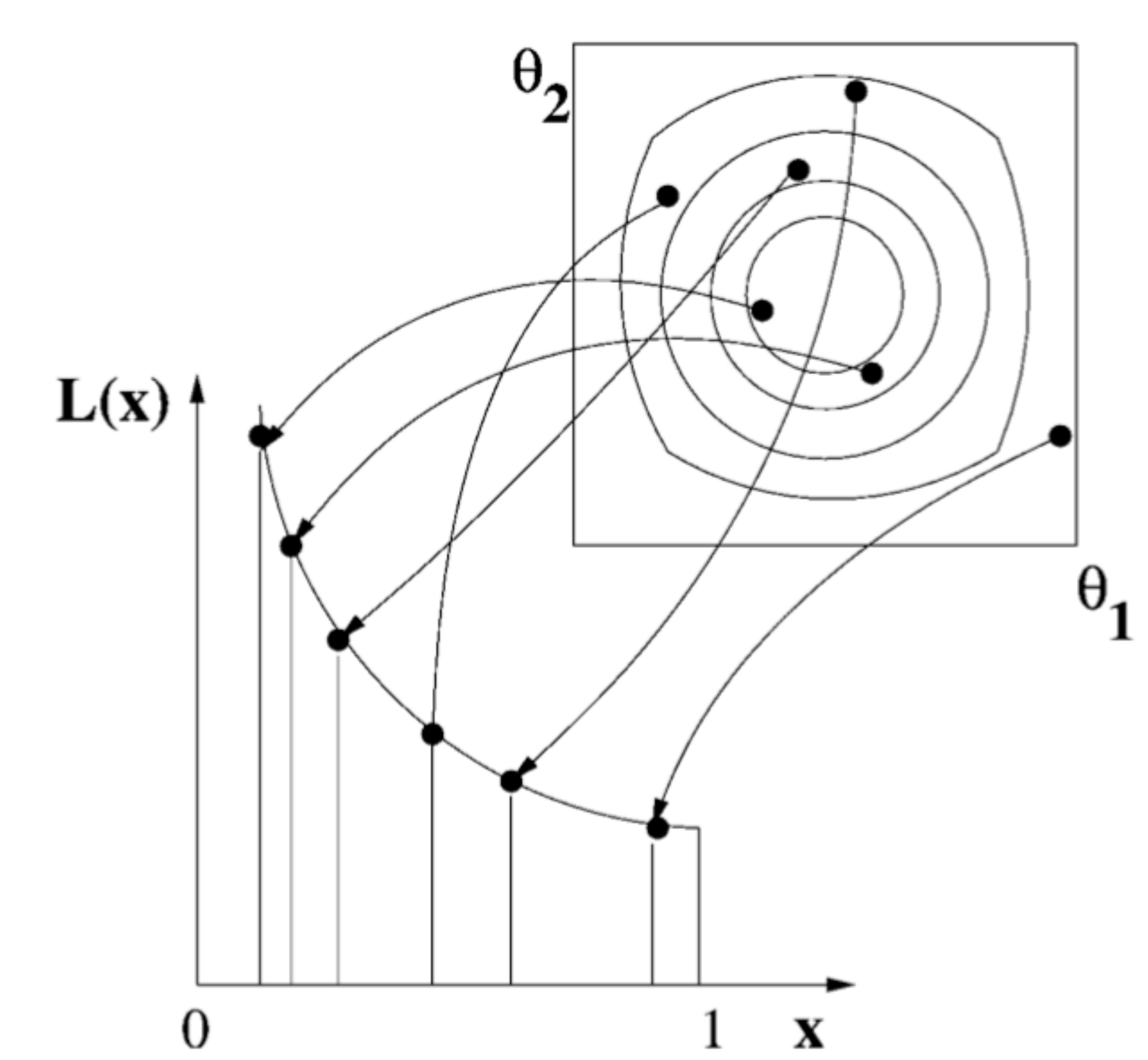}
\caption{The Nested Sampling algorithm. From \cite{MPL06}}
\label{Errors}     
\end{figure}

\section{Summary}

We have explored the use of (principally) Bayesian methods for parameter estimation and model selection, and looked at some common numerical methods for performing these tasks.  We have also shown how it is possible to work out the (minimum) errors on parameters, and on model probabilities, in advance of doing an experiment, to see if it is worthwhile.  These are the common Fisher matrix approach for parameter estimation, and the expected evidence for model selection, which requires no more than the Fisher matrix to compute.  

\section{Exercises}

\noindent 1. A Bayesian Exercise.  This is a variant of the famous {\em Monty Hall} problem, named after a game show host.    For this exercise, you must formulate the problem in a fully Bayesian way.

You are participating in a game show where there are three small doors presented to you.  You are told that behind two of them there is a bottle of fine italian wine, and behind the third is a can of the famous scottish soft drink, Irn Bru.  You will win one of them, and naturally you prefer the Irn Bru.  The game works as follows:  

\noindent a) You point to, {\em but do not open}, one of the doors.\\
b) The game show host  opens one of the other doors, revealing a bottle of wine.\\
c) You now either open the door which you originally selected, or switch to the third door.  Which should you open? 

\noindent 2. This is a model selection problem, and will get you thinking about suitable priors.   You draw lottery ticket number 1475567, and it wins.  What is the probability that the lottery was rigged, so that the winning ticket was predetermined?

\noindent 3. An exercise in linear algebra\\
a)  Prove that $(\C^{-1}),_\alpha = -\C^{-1}\C,_\alpha \C^{-1}$\\
\noindent b)  Prove that $(\ln \C),_\alpha = \C^{-1} \C,_\alpha$.\\
\noindent c)  Prove that $\ln\det \C = \Tr\ln \C$.

\noindent Hints:\\
\noindent a) What is $\C\C^{-1}$?\\
\noindent b) For matrices, $\exp(\A) \equiv \II + \A + \A^2/2! +
\ldots$.  $\ln$ is the inverse of $\exp$.\\
\noindent c) If you rotate coordinate axes (in parameter space inFevidence
this case), the determinant and trace don't change.
%

\noindent 4. A survey is proposed, to determine the mean number
density of a certain type of astronomical object, whose positions
are random. The survey measures the number of objects in each of
$N$ independent cells of the same size and shape, such that the mean number per cell, $\bar
n$, is $\gg 1$.  If the cells are observed to have $n_i$ objects
in them ($i=1,\ldots N$), show that the Fisher matrix (a scalar in
this case) for $\bar n$ is
\begin{equation}
\F = {N\over 2\bar n^2} + {N\over \bar n}.
\end{equation}
Where is most of the
information coming from, the dependence of \vmu\ on $\bar n$, or
\C? \noindent

\noindent 5. Binomial drinking - an interesting paradox.  
Precursor question.  If we have an experiment with some discrete outcomes $n=1,\ldots,\infty$, each occurring with probability $P_n$.  Argue that the probability of the sum of two drawings $z=m+n$ from the distribution is 
\[
p_z(z) = \sum_{n=1}^{z-1} P_n P_{z-n}.
\]

Every minute, as the second hand on a large clock on the wall reaches the top, I think about drinking a bottle of Irn Bru.  I drink one with a probability $p$.  Show that the probability of the next drink being taken at the $M^{th}$ opportunity is
\[
P_M = p q^{M-1}
\]
where $q=1-p$.

Show that the expectation value of the number of time steps between outbursts is $\bar M= 1/p$.  (Hint: expand $(1-q)^{-2}$ in a Taylor expansion).

Now, my friend comes in at a random time (not necessarily as the second hand reaches the top of the clock).  Argue that the probability that I took my last drink $m$ time steps previously was  
\[
P'_m = p q^{m-1}.
\]

Show that the probability for the variable $S=M+m$ ($S\ge 2$) is
\[
P_S = \sum_{M=1}^{S-1} p^2 q^{S-2}. 
\]
By expanding a different power of $1-q$, show that the expectation value of $S$ is
\[
\bar S = \frac{2}{p},
\]
and hence that the average time between the last output and the next one is
\[
\bar t = \frac{2}{p}-1.
\]
For $p=0.1$, $\bar t=19$, compared with 10 for the mean time between drinks.  How is the paradox resolved?



\section{Solutions to selected exercises}

1.  The aim here is to do a Bayesian calculation, not necessarily to find the easiest or best solution. 
Let the doors be labelled $a,b,c$, where $a$ is the door you choose
initially, and $b$ is the door which is opened.  Many of the
probabilities below should be interpreted as `given that you have
chosen $a$', but we won't write this explicitly.

Let $p(a)$ = probability that $a$ leads to the Irn Bru. $p(b)$, $p(c)$ similarly.

Let $B$ be the event that door $b$ gets opened and leads to a bottle of wine.

What you want is the probability that $a$ leads to the Irn Bru, given that $b$ is opened and leads to a bottle of wine.  
i.e. the aim is to calculate
$$
p(a|B).
$$
We can use Bayes' theorem for this:
$$
p(a|B) = {p(a,B)\over p(B)} = {p(B|a)p(a)\over p(B)}
$$
Now, clearly $p(a)=p(b)=p(c) = 1/3$ (all doors are equally likely,
before any experiment is done).

$p(B|a)$ = probability that door $b$ is opened, given that $a$ leads
to the Irn Bru.  

Since the host could have opened either door $b$ or $c$, since they
both lead to wine, we have
$$
p(B|a)={1\over 2}:
$$

What about $p(B)$? It is the sum of all the joint probabilities:
$$
p(B) = p(B,a)+p(B,b)+p(B,c) = p(B|a)p(a) + p(B|b)p(b) + p(B|c)p(c),
$$
each of which we can calculate.  $p(a)=p(b)=p(c)=1/3$, as before,
and $p(B|a)=1/2$ as before.  Now
$$
p(B|b)=0:
$$
The host will not open $b$ since it leads to the Irn Bru in this case.

$p(B|c)$ is the most interesting.  Given that you have chosen $a$
(remember this is implicit throughout), then if $c$ leads to the
Irn Bru, then the host {\it must} open door $b$,
i.e.
$$
p(B|c) = 1
$$
So the probability that your original choice $a$ leads to the
Irn Bru is
\begin{eqnarray}
p(a|B) &=& {p(B|a)p(a)\over p(B|a)p(a) + p(B|b)p(b) +
p(B|c)p(c)}\\\nonumber &=& {{1\over 2}{1\over 3}\over {1\over
2}{1\over 3}+\left(0\times{1\over 3}\right)+\left(1\times{1\over
3}\right)}\\\nonumber &=& {1\over 3}
\end{eqnarray}
So you would double your chances (from 1/3 to 2/3) if you switch to
the other door.

PS There is a really easy way to do this by lateral thinking.  After the host has opened a door, there is one remaining door with Irn Bru, and one door with wine.  Hence if you swap, you will either change success into failure, or failure into success.  Since your original chances of success were only 1/3, you improve this to 2/3 if you swap.  Easy.

\noindent 3. (a) Prove that $(\C^{-1}),_\alpha = -\C^{-1}\C,_\alpha \C^{-1}$

Since $\C\C^{-1}=\II$, its derivative is zero.  Hence
$\C(\C^{-1})_{,\alpha} + \C_{,\alpha}\C^{-1} = 0$. Result follows after
premultiplication by $\C^{-1}$.

\noindent (b) Prove that $(\ln \C)_{,\alpha} = \C^{-1} \C_{,\alpha}$.

Let $\A = \ln\C$, so
$$
\C=\exp\A = \II+\A+{\A^2\over 2!}+\ldots = \sum_{n=0}^\infty
{\A^n\over n!}
$$
Hence
$$
\C_{,\alpha} = \sum_{n=1}^\infty \,{\A^{n-1}\over (n-1)!}\,\,\A_{,\alpha} =
\sum_{m=0}^\infty\, {\A^m\over m!}(\ln\C)_{,\alpha} = \C (\ln\C)_{,\alpha}
$$
and result follows.

\noindent (c) Prove that $\ln\det \C = \Tr\ln \C$.

$\C$ is a symmetric matrix and therefore be diagonalised.  In the
new basis (where we have a new set of parameters which are linear
combinations of the old ones), the diagonal components of $\C$ are
strictly positive (they are variances of the new parameters).

Since the trace and determinant are unchanged by the
diagonalisation, we can prove the result in the rotated system. If
$\C$ is diagonal, then $\ln\C$ is diagonal, with components
$\ln\C_{11},\ \ln\C_{22}, \ldots$.  So ${\rm Tr}\ln\C =
\sum_{n}\ln\C_{nn}$.  Since $\det\C = \prod_n \C_{nn}$,
$$
\ln\det\C = \sum_n \ln\C_{nn} = {\rm Tr}\ln\C.
$$
This proof is rigorous, but there may be a neater solution without
diagonalisation.
%

\noindent 4. For $\bar n\gg 1$, we can approximate the Poisson
distribution by a gaussian, with $\langle n_i \rangle = \bar n$,
and $\sigma_i^2 = \bar n$.  Hence $\vmu^T = \bar n(1,1,\ldots 1)$
and $\C = \bar n\, {\rm diag}(1,1,\ldots,1)$.  Hence $\C_{,1} = {\rm diag}(1,1, \ldots, 1)$, and $M_{ij}$ is an $N
\times N$ matrix filled with 2s.  The result
\begin{equation}
\F = {N\over 2\bar n^2} + {N\over \bar n}
\end{equation}
follows.  The first term arises from $\C_{,1}$, and the second
from $\vmu_{,1}$.  The second dominates since $\bar n \gg 1$.


\begin{thebibliography}{99.}

\bibitem{Amendola}Amendola L., Kunz M., Sapone D., JCAP, 04, 013 (2008)

\bibitem{Beltran05}
Beltran M., Garcia-Bellido J., Lesgourgues J., Liddle A. R., Slosar A., 2005, Phys, Rev., D71, 063532

\bibitem{Duan}Duan et al., Phys. Lett. B195, 216 (1987)

\bibitem{Feroz}Feroz F., Hobson M., Bridges M., astroph/0809.3437 (2008)

\bibitem{Fisher} Fisher R. A., J. Roy. Stat. Soc., 98, 39
(1935)

\bibitem{Gilks} Gilks W.R, Richardson S., Spiegelhalter D.J., {\em
Markov chain Monte Carlo in practice}, Chapman \& Hall, London
(1996)

\bibitem{Hajian06}Hajian A., astroph/0608679 (2006)

\bibitem{Hamilton05}Hamilton A.J.S., astroph/0503603 (2005)

\bibitem{Heavens07} Heavens A.F.,  Kitching T., Verde L., MNRAS, 380, 1029 (2007).

\bibitem{Hobson}
Hobson M.P., Bridle S.L., Lahav O., 2002, MNRAS, 335, 377

\bibitem{Jeffreys61}
Jeffreys H., 1961, Theory of Probability, Oxford University Press
(Oxford, UK)

\bibitem{KS} Kendall M. G., Stuart A., {\it The Advanced Theory of Statistics,
Volume II}, Griffin, London (1969)

\bibitem{KK} Kenney, J. F. \& Keeping, E. S., {\it Mathematics of Statistics, Part II},
2nd ed. (Van Nostrand, New York) (1951)

\bibitem{Kosowsky02}Kosowsky A., Milosavljevic M., Jimenez R., Phys. Rev. D66, 3007 (2002).

\bibitem{Lazarides}
Lazarides, G., Ruiz de Austri, R., Trotta, R., Phys. Rev. D70, 123527

\bibitem{Lewis} Lewis A., Bridle S., PRD, 66, 103511 (2002)

\bibitem{Liddle07} Liddle A., MNRAS, 377, 74 (2007).


\bibitem{MPL06}
Mukherjee P., Parkinson D., Liddle A.R., 2006, ApJ, 638, L51


 


\bibitem{NumRec} Press W., et al, {\em Numerical Recipes in Fortran},
CUP, Cambridge, U.K. (1992)


\bibitem{Serra07} Serra P., Heavens A., Melchiorri A.,  MNRAS 379, 169 (2007)

\bibitem{Skilling}
Skilling J., 2004, avaliable at\\
http://www.inference.phy.cam.ac.uk/bayesys


\bibitem{TTH} Tegmark M., Taylor, A.N., Heavens A.F., ApJ, 480, 22 (1997) (TTH)


\bibitem{Trotta07}
Trotta R., 2007, astroph/0703063

\bibitem{VerdeNotes}Verde L., astroph/0712.3028 (2007)

\bibitem{Verde} Verde L., et al, ApJS, 148, 195 (2003)


\end{thebibliography}
\end{document}